\documentclass[12pt]{article}
\newcommand{\be}{\begin{equation}}
\newcommand{\bea}{\begin{eqnarray}}
\newcommand{\eea}{\end{eqnarray}}
\newcommand{\ba}{\begin{array}}
\newcommand{\ea}{\end{array}}
\newcommand{\ee}{\end{equation}}

\expandafter\ifx\csname mathbbm\endcsname\relax

\else

\fi
\def\l{\label}

\begin{document}
\begin{titlepage}
\hfill
\vbox{
    \halign{#\hfil         \cr
           IPM/P-2003/060 \cr
           hep-th/0309097  \cr
           } 
      }  
\vspace*{20mm}
\begin{center}
{\Large {\bf Glueball Superfield and Argyres Douglas Points }\\ }

\vspace*{15mm}
\vspace*{1mm}
{Mohammad A. Ganjali}
 \\
\vspace*{1cm}

{\it Institute for Studies in Theoretical Physics
and Mathematics (IPM)\\
 \vspace{3mm}
 Department of Physics, Sharif University of Technology\\
P.O. Box 11365-9161, Tehran, Iran}\\
\vspace*{.5cm}
E-mail: Ganjali@mehr.sharif.edu\\

 \vspace*{1cm}
\end{center}

\begin{abstract}
In this paper, we study ${\cal N}=1$ super-symmetric $SO(N)$ gauge
theory in Argyres-Douglas points by using the factorization
equation of the ${\cal N}=2$ theory. We suppose that all
monopoles become massive in the system and obtain a tree level
superpotential. Then, we obtain general Picard-fuchs equations
for glueball superfields which are hypergeometric equations having
regular singular points corresponding to Argyres-Douglas points.
Furthermore, we study the solutions of these differential
equations and calculate the effective superpotential. Finally, we
study scaling behavior of the chiral operators and coupling
constants around the AD points.

\end{abstract}

\end{titlepage}

\section{Introduction}

$\;\;\;\;$Recently, Dijkraaf and
 Vafa \cite {Dijkgraaf:2002dh,Dijkgraaf:2002vw,Dijkgraaf:2002fc} have
understood the quantum dynamics of a wide class of ${\cal N}=1$
supersymmetric gauge theory by studying an auxiliary matrix model
where they conjectured that the effective superpotentials in
${\cal N}=1$ SQCD's theories have a correspondence with the free
energy of matrix model. The nonperturbative effects of gauge
theory can be also obtained from the planar diagrams of matrix
model. At first, these correspondence had been proved by
considering the geometric transition and topological string theory
 \cite {Vafa:2000wi,Cachazo:2001jy,Cachazo:2002pr} and Cachazo,
Douglas, Seiberg and Witten's efforts were caused them to be able
to prove this correspondence by the field theory concepts without
using string theory
 \cite {Cachazo:2002ry,Seiberg:2002jq,Cachazo:2002zk,Cachazo:2003yc}
. In fact, they applied a technology which was based on the
anomalous Ward identity of generalized Konishi anomaly \cite
{Konishi:1983hf,Konishi:1985tu} . Moreover, they used the
factorization equation \cite {Cachazo:2001jy,Cachazo:2002pr} which
relates the ${\cal N}=2$ Seiberg-Witten curves \cite
{Seiberg:1994rs,Seiberg:1994aj} to reduced ${\cal N}=1$ curve in
the presence of massless monopoles. Given the tree-level
superpotential ${\cal W'}(x)$ the factorization equation
completely determines all the parameters of polynomials which
exist in the factorization equation and so the vacuum structure of
theory can be identified.Furthermore, one may calculate the
monopoles condensate which show at generic points in parameter
space this quantity can be non zero and generates mass gapp and
confinement in the system.However, Using this geometric picture,
when all monopoles become massive in the system, Physics is the
same as ${\cal N}=2$ theory and One may derive  a system of
ordinary differential equations, Picard-fuchs equations, for
glueball superfields and calculates the exact effective
superpotential.

One can combine this method with matrix model and study the non-
perturbative effects of ${\cal N}=1$ theory, as well.

Furthermore, these methods give us ability to study other
interesting phenomena such as conformal behavior of supersymmetric
gauge theory in points called $Argyres-Douglas$ points \cite
{Argyres:1995jj,Seiberg:1994rs,Klemm:1994qs,Argyres:1994xh}. In
fact, one can break the ${\cal N}=2$ supersymmetry with a
perturbed superpotential but in points where monopoles become
massless, one can reproduce ${\cal N}=1$ supersymmetry. Following
this method, one can find where reduced {${\cal N}=1$} theory has
{\it conformal} invariance is this point in \cite
{Argyres:1995jj,Argyres:1995xn,Eguchi:1996vu,Eguchi:1996ds}.

Even more interestingly it had been noted in \cite
{Ferrari:2002jp,Ferrari:2002kq} that for an ${\cal N}=1$ gauge
theory with cubic superpotential when the gauge group is unbroken,
there are critical values of superpotential couplings where the
effective superpotential is non-analytic and so the large $N$
expansion of such an effective superpotetial is singular too. But
these singularities can be removed by the {\it double scaling
limit}. Furthermore, these double scaling limits define an ${\cal
N}=1$ four dimensional non-critical string theory \cite
{Ferrari:2002ad,Ferrari:2002gy} .

These proposal along with the machinary that was introduced in
\cite
{Cachazo:2002ry,Seiberg:2002jq,Cachazo:2002zk,Cachazo:2003yc}, was
performed in  \cite {Eguchi:2003wv} and \cite {Bertoldi:2003ab}
for $U(N)$ gauge group. Now, we would like to have a survey of
this proposal for other classical gauge groups $SO(2N)$ and
SO$(2N+1)$. Therefore, at absence of massless monopoles, we try
to find the general Picard-Fuchs equations for glueball
superfield without flavors then, we'll solve these equations.
These differential equations are the Picard-Fuchs equations for
the periods of the {\it memomorphic} one-form ${\cal ydx}$ on the
spectral curve. The superpotential is also obtained and the
conformal behavior of theory will be studied by taking the IR
limit respectively.

This paper is organized as follows. In section $2$ we review the
CDSW's method for calculating the chiral operators and study the
${\cal N}=1$ U(N) gauge theory around the AD points. In section
$3$ we study the factorization equation of $SO(N)$ gauge group and
then we focus on AD points and obtain ${\cal N}=1$ effective
superpotential in terms of chiral superfield $\Phi$.

In section $4$ we find the Picard-Fuchs equations for glueball
superfields for gauge groups U(N) and SO(N) which are in general
hypergeometric equations. Then we find the solutions of these
equations for $SO(2N)$ and $SO(2N+1)$ respectively. In the last
section we obtain the superpotential of $SO(N)$ theory and study
the scaling behavior of chiral fields around the AD points.

\section{Review of U(N) theory}

$\;\;\;$ The dynamic of ${\cal N}=1$ $U(N)$ gauge theories with
superpotential ${\cal W}(\Phi)$ can be studied as a perturbation
of the ${\cal N}=2$ strongly coupled gauge theory with ${\cal
W}=0$. The low energy group is $U(1)^n$ and $N-n$ monopoles of
the ${\cal N}=2$ theory are massless. The Seiberg-Witten curve has
the following factorization at this points \cite
{Cachazo:2001jy,Cachazo:2002pr},
   \bea\label {f1}
       y ^{2}=P_{N}^{2}(x)-4\Lambda^{2N}=F_{2n}H^{2}_{N-n}(x),
       \eea
where the polynomyal in the r.h.s has simple roots. In \cite
{Cachazo:2001jy} Cachazo,Intriligatore and Vafa showed that
    \bea\label {f2}
         F_{2n}(x)=\frac{1}{g^{2}_{n}}{\cal W }'(x)^{2}+{\it
         f}_{n-1}(x),
    \eea
 where ${\cal W}(x)$ is the superpotential for the reduced
${\cal N}=1$ theory and is a polynomial of degree $n$. From this
factorization, the gauge group $U(N)$ breaks to $U(1)^{n}$ and so
$N-n$  monopoles become massless. The CDSW's method for the
calculation of chiral operators is as below \cite
{Cachazo:2002ry,Seiberg:2002jq,Cachazo:2002zk,Cachazo:2003yc}.
     \bea\label {f3}
          T(x)&=&\langle
          Tr\frac{1}{x-\Phi}\rangle ,\\ R(x)&=&-\frac{1}{32\pi^{2}}\langle
          Tr\frac{W_{\alpha}W^{\alpha}}{x-\Phi}\rangle,
     \eea
 where
$W^{\alpha}$  is the field strength chiral field. In terms of SW
curve
    \bea \label {f4}
         T(x)&=&\frac{P_ {N}'}{y_{{\cal N}=2}(x)},\\
         R(x)&=&\frac{1}{2} ({\cal W}'(x)-y_{{\cal N}=1}(x)).
    \eea
The expectation value of chiral fields are also given by $U_{r}$
and $S_{r}$ where
    \bea\label {f5}
         U_{r}&\equiv& \langle Tr\Phi^{r}\rangle=\oint x^{r}T(x)dx,\\
         S_{r}&\equiv&\langle Tr\Phi^{r}W_{\alpha}W^{\alpha}\rangle=\oint
         x^{r}T(x)d x.
    \eea
 Finally, when $U(N)$ theory has been broken to
$\Pi_{k=1}^n U(N_k)$, the effective superpotential is
    \bea\label {f6}
         {\cal W}_{eff}(S)=\sum_{k=1} ^{n}N_k\frac{\partial{{\cal
         F}}}{\partial{S_k}}+2\pi i \tau_0\sum_{k=1}^{n}S_k+2\pi
         i\sum_{k=2}^{n}b_k S_k,
    \eea
where ${\cal F}$ is prepotential. Glueball superfields and
prepotential are generally given by
    \bea \label {f7}
          S_{k}=\frac{1}{2\pi i}\oint_{A_{k}}\textit{y d x},\hspace{1cm}
         \frac{\partial{{\cal F}}}{\partial{S_k}}=\oint_{B_
         k}{ydx},
    \eea
and
    \bea\label {f8}
         N_{i}=\frac{1}{2\pi i}\oint_{A_{k}}T(x)dx,\hspace{2cm}\\
         \tau_{0}=\frac{1}{2\pi i}\int_{B_{1}}\overline{T},
         \hspace{.5cm}b_{k}=-\frac{1}{2\pi
         i}\int_{B_{k}}\overline{T}-\tau_{0}\l {h8},
    \eea
where the $A_k$'s are the closed circles around the branch cuts of
the spectral curve and the $B_k$'s are the non-compact cycles
connecting the points at infinity on the two sheets of the
spectral curve passing through the $A_k-$th branch cut. The
intersection pairs of these cycles are
    \bea \label {f9}
         A_i\bigcap A_j=B_i\bigcap
         B_j=0,\hspace{.5cm}A_i\bigcap B_j=\delta_{ij}.
    \eea
Let us introduce  $C_k$ cycles for the future use
    \bea
         C_k=B_{k+1}-B_k,\hspace{.5cm}k=1,...,n-1,
    \eea
We also introduce a small cycle $A_0$ around the origin $x=0$.

Now, we recall that Argyres-Douglas points \cite
{Argyres:1995jj,Seiberg:1994rs,Klemm:1994qs,Argyres:1994xh} occur
where the ${\cal N}=2$ gauge theory exhibits the ${\cal N}=2$
superconformal symmetry\cite
{Argyres:1995jj,Argyres:1995xn,Eguchi:1996vu,Eguchi:1996ds}. In
fact, the vanishing cycles which have non-trivial intersections,
imply that the low-energy ${\cal N}=2$ theory has massless
solitons with both electrical and magnetical charge under the same
$U(1)$ factor\cite {Argyres:1995jj}. These points in moduli space
correspond to higher order singularities and are simply obtained
by adjusting the moduli parameters of the characteristic
polynomial $P_{N}(x)$. For example consider the following
Seiberg-Witten curve for $SU(N)$
    \bea
         y^2=P_N(x)^2-4\Lambda^{2N}=(x^N-u)^2-4\Lambda^{2N},\cr
         P_N(x)=Det(x.I-\Phi(x))=x^N-\sum_{i=0}^{N-1}u_i x^i.
    \eea
Then AD points are the zeros of discriminants of this curve which
are
    \bea
    s_i=0, \;\;\;(i=1,...,N-1),\;\;\;s_0=\pm 2\Lambda^N.
    \eea
This curve has a $Z_N$ symmetry where
    \bea \label {f10}
         (x,y)\rightarrow(\textit{e}^\frac{2\pi i}{N}x,y).
    \eea

Being broken of gauge group from $U(N)$ into $U(1)^{N}$ dose in
the presence of this superpotential
    \bea \label {f11}
         {\cal W}(\Phi)=g_{N}(\frac{1}{N+1}\phi^{N+1}-u\phi),\;\;\;\;N\geq3.
    \eea
Now, one can find the Picard-Fuchs equation as \cite
{Bertoldi:2003ab}
    \bea \label {f12}
         [\partial_{u}^2+(\frac{N-2}{N})\frac{1}{u^2-4\Lambda^{2N}}\partial_u-
         (\frac{N^2-1}{N^2})\frac{1}{u^2-4\Lambda^{2N}}]S_k=0,
    \eea
that are a hypergeometric equations. According to these
consequences
    \bea \label {f13}
         {\cal W}_{eff}(S)=2\pi i\sum_{k=1}^{N}b_{k}S_{k},
    \eea
and the effective superpotential of the $U(N)$ theory is \cite
{Bertoldi:2003ab}
    \bea \label {f14}
         {\cal W}_{eff}(s)=-2\pi i \frac{N}{e^{2\pi
         i/{N}}-1}S_1(u,\Lambda^{2N}),
    \eea
where
    \bea \label {f15}
         S_1(u,\Lambda^{2N})=-\frac{2\Lambda^{2N}}{N}\textit{e}^{2\pi
         i/{N}}u^{-1+\frac{1}{N}}F(\frac{1}{2}-\frac{1}{2N},1-\frac{1}{2N},2,
         \frac{4\Lambda^{2N}}{u^2}).
    \eea
and the function {\it F} will be defined in section 4. As it would
be shown in \cite {Bertoldi:2003ab}, this effective superpotential
has a non trivial behavior in large $N$ limit but it can be
removed by using Double-scaling limit.

Besides, in \cite {Eguchi:2003wv}, the scaling behavior of this
theory in the IR limit where $\Lambda\rightarrow\infty$  was
studied and the scaling dimensions of chiral operators were
obtained
    \be \label {g16}
         U_r=0,\;\;\;\;\;\;S_r=0,\;\;\;\;\;\;{\rm for\; all\; r}.
    \ee
These are consistent with scaling invariance. In addition, if we
consider small perturbation around the AD points as
     \bea \label {f17}
         {\cal W}'(x)=P_N(x)=x^N-2\Lambda^N-\sum_{m=0}^{N-1}g_mx^m\Lambda^{N-m},
     \eea
then the scaling dimension for chiral operators and coupling
constants are
    \bea \label {f18}
         \Delta(g_m)=\frac{2(N-m)}{N+2},\hspace{.5cm}\Delta(U_r)=\frac{N+2r}{N+2},
         \hspace{.5cm}\Delta(S_r)=\frac{N+2(r+1)}{N+2}.
    \eea
It is noticeable that
    \bea \label {f19}
         \Delta(g_m)\leq 1\Leftrightarrow m\geq\frac{N}{2}-1\;\;\;  ({\rm
         even\; N})\; {\rm or} \;\;\;\;\; m\geq[\frac{N}{2}]\;\;\;
         ({\rm odd\; N}),
    \eea
that corresponds to a coupling constant in the ${\cal N}=2$
superconformal field theory in $4$ dimension \cite
{Argyres:1995jj,Argyres:1995xn,Eguchi:1996vu,Eguchi:1996ds}.

$\;\;\;\;\;\;$\section{The factorization and spectral curve of
SO(N)}

In this section, we use the factorization of Seiberg-Witten curve
to obtain ${\cal N}=1$ spectral curve for gauge groupe $SO(N)$.

At first, we demonstrate for this gauge group
    \bea \label {f20}
         y^2=P_{2N}^2(x)-\Lambda^{2
         \widehat{h}}x^{2l}=x^2[(T_{2N-1}(x))^2-\Lambda^{2\widehat{h}}x^{2l-2}],
    \eea
 where
    \bea \label {f21}
         SO(2N):\widehat{h}=2N-2,\hspace{.5cm}l=2,\cr
         SO(2N+1):\widehat{h}=2N-1,\hspace{.5cm}l=1.
    \eea
The factorization equation for gauge group $SO(N)$ is \cite
{Ahn:1997wh, Edelstein:2001mw,Fuji:2002vv}
    \bea \label {h21}
         y^2=x^2(H_{2N-2n-2}(x))^2F_{4n+2}=x^2(H_{2N-2n-2}(x))^2({\cal W}'^2+f_{2n}),
    \eea
so, the reduced ${\cal N}=1$ curve is
    \bea
         y_{{\cal N}=1}=F_{4n+2}={\cal W}'^2+f_{2n}.
    \eea
Then, we follow
    \bea \label {f22}
         y^2&=&x^2[(T_{2N-1}(x))-\Lambda^{\widehat{h}}x^{l-1}]
         [(T_{2N-1}(x))+\Lambda^{\widehat{h}}x^{l-1}]\cr
         &=&x^2(H_{2N-2n-2}(x))^2F_{4n+2}.
    \eea
Just as $2N-2n-2$ double roots occur in $(H_{2N-2n-2}(x))^2$, then
$2N-2n-2$  monopoles become massless. Since
$(T_{2N-1}(x)-\Lambda^{\widehat{h}}x^{l-1})$ and
$(T_{2N-1}(x)+\Lambda^{\widehat{h}}x^{l-1})$ can not share any
zeroes, we can classify the solutions of (\ref {f22}) based on how
to divide the zeroes of $(H_{2N-2n-2}(x))^2$ into these two
factors where

\begin{enumerate}
    \item  $s_{+}$: All the zeroes of $(H_{2N-2n-2}(x))^2$ are  those of
$T_{2N-1}(x)+\Lambda^{\widehat{h}}x^{l-1}$

    \item  $s_{-}$: All the zeroes of
$(H_{2N-2n-2}(x))^2$ are those of
$T_{2N-1}(x)-\Lambda^{\widehat{h}}x^{l-1}$

    \item $s_{+}s_{-}\neq 0$: The case other than(i) and (ii).
\end{enumerate}
Also
    \bea
         s_{+}+s_{-}=2N-2n-2.
    \eea

Let us  consider the case ($1$). One might obtain
   \bea \label {f23}
         T_{2N-1}(x)+\Lambda^{\widehat{h}}x^{l-1}&=&(H_{2N-2n-2}(x))^2K_{4n-2N+3}\cr
         T_{2N-1}(x)-\Lambda^{\widehat{h}}x^{l-1}&=&(H_{2N-2n-2}(x))^2K_{4n-2N+3}
         -2\Lambda^{\widehat{h}}x^{l-1},
    \eea
then
    \bea \label {f24}
         {\cal W}'(x)^2&=&[H_{2N-2n-2}(x)K_{4n-2N+3}]^2\\
         f_{2n}(x)&=&-2\Lambda^{\widehat{h}}x^{l}K_{4n-2N+2}.
    \eea
Therefore, because the functions ${\cal W}$ and $H$ and $K$ are
functions of $x^2$, we can write the complete set of the their
zeroes as
    \bea \label {f25}
         {\cal W}'(x)&=&\prod_{i=1}^{m}(x^2-a_{i}^2),\\
         H_{2N-2n-2}(x)&=&\prod_{i=1}^{N-n-1}(x^2-p_{i}^2),\\
         K_{4n-2N+2}(x)&=&\prod_{j=1}^{2n-N+1}(x^2-q_{j}^2),
    \eea
where
    \bea \label {f26}
         \{p_1,...,p_{N-n-1}\}\bigcup\{q_1,...,q_{2n-N+1}\}=\{a_1,...,a_{m}\}.
    \eea
Following the case of $U(N)$ gauge theory in \cite
{deBoer:1997ap,Eguchi:2003wv}, and according to
\cite{Ahn:1997wh}, one might obtain this expression for monopole
condensation for $SO(N)$
    \bea \label {f27}
         \langle{\widetilde{M_{i}}M_i}\rangle&=&const\times[x^{2l-1}\times y_{{\cal N}=1}]_{(x=p_i)}\\
         &=&const\times\prod_{j=1}^{2n-N+1}p_i^{\frac{5l-2}{2}}(p_i^2-q_j^2)^{\frac{1}{2}},
    \eea
where $M_{i},\widetilde{M_{i}}$ are the scalar components of i-th
monopole hypermultiplet and by (\ref {f27}) the vev of
$i=1,...,2N-2n-2$ monopoles are all non zero and generate the mass
gap and confinement in the system. Similar arguments can be given
for the case $(2)$ and one may easily derive
    \bea \label {f28}
         {\cal W}'(x)^2&=&[(H_{2N-2n-2}(x))^2(K_{4n-2N+3})]^2\\
         f_{2n}(x)&=&2\Lambda^{\widehat{h}}x^{l}K_{4n-2N+2}.
    \eea
The case $(3)$ is more complicated than the others  and  it is not
necessary to consider it for this paper.

Now, by considering
    \bea \label {f29}
         y^{2}=(x^{2N}-u x^2)^2 -\Lambda^{2\widehat{h}}x^{2l},
    \eea
and writing the discriminant  of this curve, one finds that the
above curve has higher singularities and one can calculate the
Argyres-Douglas point. In particular, for gauge group $SO(2N)$
    \bea \label {f30}
         y^2=x^4(x^{2N-2}-u+\Lambda^{2N-2})(x^{2N-2}-u-\Lambda^{2N-2}),
         \cr
         u=\pm{\Lambda^{2N-2}},\hspace{4cm}
    \eea
and for gauge group $SO(2N+1)$, for calculating the discriminant,
one must solve this algebraic equation
    \bea \label {h30}
         x^2[(x^{2N-1}-u x)^2-\Lambda ^{4N-2}]=0.
    \eea
As an illustration, let us consider the $SO(5)$ case. The
Argyres-douglas points of this curve are
    \bea \label {k30}
         u_1&=& \frac{3}{\sqrt[3]{4}}e^{\frac{-2\pi i
         s}{3}}\Lambda^{\frac{4N-2}{3}},\;\;\;\; s=1,2,3,\\ \cr
         u_2&=& 3e^{\frac{-2\pi i
         s}{3}}\Lambda^{\frac{4N-2}{3}},\;\;\;\; s=1,2,3.
    \eea
In these points, AD points, the effective superpotential has
non-analytic behavior and has conformal invariance.

The curve (\ref {f29}) corresponds to the the case where non of
the monopoles are masslesss. On the other hand, if we choose
$n=N-1$ where non of the monopoles become massless,
$H_{2N-2n-2}=H_0=1$ and
    \bea
         {\cal W}'(x)=K_{2N-1},
    \eea
then
    \bea
         {\cal W}'(x)&=&x^{2N-1}-ux,\\
         f_{2n}&=&-\Lambda^{2\widehat{h}}x^{2l},
    \eea
and finally
    \bea \label {f31}
         {\cal W}(x)=\frac{1}{2N}x^{2N}-\frac{u}{2}x^2.
    \eea

$\;\;\;\;\;\;$\section{ Picard-Fuchs equations for glueball
superfield}

As it was mentioned, to find the glueball superfield in the
geometric picture, one must calculate the following integral
    \bea
         S_{k}=\frac{1}{2\pi i}\oint_{A_{k}}y d x.
    \eea
Now, we follow the method that was introduced in \cite
{Ito:1996sq}, for obtaining the Picard-Fuchs equations for
classical gauge group $U(n)$ and $SO(n)$. If we consider the case
where all the monopoles are massive in the theory then the  ${\cal
N}=1$ curve is the same as ${\cal N}=2$ curve and one may use the
following form of Seiberg-witten curve to calculate the glueball
superfield,
    \bea
         y^2=P^2(x)-\Lambda^{2\widehat{h}}x^{2l},
    \eea
where $\widehat{h}$ is the dual coexter number of the Lie gauge
group and
    \bea
         P(x)=x^n-\sum^n_{i=2}u_ix^{n-i},
    \eea
with $n=r+1,i=2,3,...,r+1$ for $A_r$ series and
$n=2r,i=2,4,...,2r$ for $B_r,D_r$ series, and $u_i$'s, are the
casimirs of the gauge groups. Also $l=n-\widehat{h}$ .

Base on the explicit form of $S_k$ and using the fact that $S_k$
is linearly independent of the casimirs, setting
$\frac{\partial}{\partial u_i}=\partial_i$ then
    \bea \label {f32}
         \partial_i{S_k}&=&-\frac{1}{2\pi i}\oint{\frac{x^{n-i}P}{y}}\\
         \partial_i\partial_j{S_k}&=&-\frac{1}{2\pi i}
         \oint{\frac{x^{2n+2l-i-j}\Lambda^{2\widehat{h}}}{y^3}}\l {h32},
    \eea
and using a direct calculation, one obtains
    \bea \label {f33}
         \frac{d}{d x}(\frac{x^mP}{y})=
         m\frac{x^{m-1}P}{y}-\frac{\Lambda^{2\widehat{h}}}{y^3}(n-l)x^{m+n+2l-1}
         +\frac{\Lambda^{2\widehat{h}}}{y^3}\sum_t(n-t-l)u_tx^{m+n+2l-t-1}.
    \eea
Comparing equations (\ref {f32}) and (\ref {h32}) with (\ref
{f33}) we can find the second order differential equations for the
$S_k$'s as follow
    \bea \label {f35}
         \pounds_{m}=-m\partial_{n-m+1}-(n-l)\partial_2\partial_{n-m+1}+
         \sum_{i}(n-i-l)u_i\partial_i\partial_{n-m+1},
    \eea
where $m=t-1$ for $A_r$ series and $m=2lt-1$ for $B_r$ and $D_r$
series and $t=1,...,r-1$. Note that in $A_r$ series for $t=1$ the
above expression does not give the correct equation. By looking at
the final step of (\ref {f35}), one might obtain the following
identity
    \bea \label {h35}
          \pounds_{0}^{A_r}=-(n-l)\partial_2\partial_{r}+
         \sum_{i}(r+1-i)u_i\partial_{i+1}\partial_{r+1}.
    \eea
For $t>r-1$ equation (\ref {f35}) does not give the second order
differential equation for $u_i$. Notice that according to equation
(\ref {f35}) we have
    \bea
         \pounds_{i,j;p,q}=\partial_i\partial_j-\partial_p\partial_q,\hspace{.5cm}i+j=p+q.
    \eea
The $r-$th equation, the {\it exceptional} equation, can be obtain
according to this linear independent identity
    \bea \label {h36}
         -(n-l)\oint_{A_k}d(\frac{x^{n+1}P}{y})+
         \sum_{i}(n-l-i)u_i \oint_{A_k}d(\frac{x^{n+1-i}P}{y})=0.
    \eea
In the expansion of the above identity there is a term as
    \bea \label {i36}
         \Upsilon=(n-l)^2\oint_{A_k}\frac{\Lambda^{4\widehat{h}}x^{4l}}{y^3}.
    \eea
We must be careful if we want to reproduce this term by partial
derivative of $S_k$. If $u_{n}$ does not be zero for $A_{r}$ or
$D_r$ series this term equals with
    \bea \label {j36}
         \Upsilon=-(n-l)^2
         \Lambda^{2\widehat{h}}\partial_{\widehat{h}}^2 S_k,
    \eea
and if $u_{2n}$ and $u_{2n-2}$ do not be zero for $B_r$ series
this term should be changed as
    \bea \label {k36}
         \Upsilon=-(n-l)^2
         \Lambda^{2\widehat{h}-1}\partial_{\widehat{h}+1}^2 S_k,
    \eea
Then
    \bea \label {f37}
         [-(n+1)(n-l)-(l+1)\sum_{i}iu_i\partial_i+\sum_{i}i^2u_i\partial_i+\cr
         \sum_{i,j}iju_iu_j\partial_i\partial_j
         +(l+1)\Lambda\partial_{\Lambda}+\Upsilon]S_k=0.
    \eea
From the fact that the $S_k$'s are homogeneous functions of degree
$n+1$, one can see that
    \bea \label {t37}
         [\Lambda\partial_{\Lambda}+\sum_{i}iu_i\partial_i]S_k=(n+1)S_k,
    \eea
therefore, the exceptional equation reads as
    \bea \label {f38}
         [(n+1)(2l-n+1)+\sum_{i}i(i-2(l+1))u_i\partial_i\cr
         +\sum_{i,j}iju_iu_j\partial_i\partial_j+\Upsilon]S_k=0.
    \eea

These exceptional equations are in general {\it hypergeometric}
equations. But, if the above conditions does not be satisfied in
an special form of geometric curve then one may use the following
identity
    \bea \label {g36}
          \Upsilon=(n-l)^2\oint_{A_k}\frac{\Lambda^{4\widehat{h}}x^{4l}}{y^3}=
          [2\widehat{h}\Lambda \partial_{\Lambda}
          -\Lambda \partial_{\Lambda}(\Lambda
          \partial_{\Lambda})].
    \eea
Neverthelse, by (\ref {t37}), one can show that this method does
not give a generic second order differential equation for gleuball
superfield. This occurs in our paper for gauge group $SO(2N+1)$
hence we focus on semi classical region where $\Upsilon$ goes to
zero. We'll study the solutions of these equations for gauge
groups $SO(2N)$ and $SO(2N+1)$ with details in the following
sections.

$\;\;\;\;\;\;$\subsection{Solution of the Picard-Fuchs equation}
In order to solve the Picard-Fuchs equations, at first, we note
that for a $U(N)$ gauge theory in which all $u_i$'s$=0$ unless
$u_N=u$ (higher AD point), the exceptional equation (\ref {f38})
reduces to equation (\ref {f12}) and the solution of this equation
is
 \bea \label {j38}
         S_k(u,\Lambda^{2N})=-\frac{2\Lambda^{2N}}{N}\textit{e}^{2\pi
         ik/{N}}u^{-1+\frac{1}{N}}F(\frac{1}{2}-\frac{1}{2N},1-\frac{1}{2N},2,
         \frac{4\Lambda^{2N}}{u^2}).
    \eea
 Now we derive solutions of differential equations for other classical gauge groups.

\subsubsection*{The case SO(2N)}

For this gauge group we have
    \bea \label {f39}
         n=2N,\hspace{.5cm}\widehat{h}=2N-2,\hspace{.5cm}l=2,
    \eea
and because of the maximal confining phase of the system we have
    \bea \label {f40}
         u_{2N-2}=u,\;\;\;\;\;{\rm other=0}.
    \eea
Therefore, we only need the exceptional equation that is
    \bea \label {f41}
         [(u^2-\Lambda^{2\widehat{h}})\partial^2_u+
         \frac{N-4}{N-1}u\partial_u+\frac{(2N+1)(5-2N)}{4(N-1)^2}]S_k=0.
    \eea
By following change of variable
   \bea \label {f42}
         z=\frac{u^2}{\Lambda^{2\widehat{h}}},
    \eea
this equation becomes a hypergeometric equation as
    \bea \label {f43}
         [z(1-z)\frac{\partial^2}{\partial
         z^2}+(\frac{1}{2}-(\frac{2N-5}{2N-2)})z)\frac{\partial}{\partial
         z}-\frac{(2N+1)(5-2N)}{16(N-1)^2}]S_k=0,
   \eea
which can be solved as \cite {Bateman1955}
    \bea \label {f44}
         S_k(z)=C_{1,k}F(a,b,c;z)+C_{2,k}z^{1-c}F(a-c+1,b-c+1;2-c;z),
    \eea
 where
    \bea \label {f45}
         a=-\frac{5-2N}{4(N-1)},\hspace{.5cm}b=-\frac{2N+1}{4(N-1)},\hspace{.5cm}c=\frac{1}{2},
    \eea
and
    \bea \label {f46}
         F(a,b,c;z)=\sum_{n=0}^{\infty}\frac{(a)_n(b)_n}{(c)_n}\frac{z^n}{n!}\hspace{.5cm}
         (a)_n=\frac{(a+n-1)!}{(a-1)!}.
    \eea
Now to fix the coefficients $C_{1,k}$ and $C_{2,k}$ one can
evaluate the $S_k$ in the semi classiacal limit
$\Lambda\rightarrow0$, where the glueball superfield vanishes.
This can be done by performing the analytic continuation of the
hypergeometric function \cite {Bateman1955}, which are defined as
power series in the disk $|z|\leq1$.

 First of all, one may show that
    \bea \label {f47}
         C_{1,k}=(-1)^{c}
         \frac{\Gamma(b)\Gamma(c-a)\Gamma(2-c)}{\Gamma(c)\Gamma(1-a)\Gamma(b-c+1)}C_{2,k}.
    \eea
After that, we rewrite the above hypergeometric function in terms
of a new variable which is dual with $\Lambda\rightarrow 0$ and we
are able to expand the solutions around the zero. Furthermore, in
$\Lambda\rightarrow 0$ the solutions of (\ref {f41}) are
asymptotic to
    \bea \label {f48}
         {u^\alpha}^{\pm} f(u),\hspace{2cm}\cr
         \alpha_{+}=\frac{5-2N}{2(N-1)},
         \alpha_{-}=\frac{2N+1}{2(N-1)},
    \eea
then, by considering the $z=\frac{\Lambda^{2\widehat{h}}}{u^2}$
and $S_k=u^{\alpha_{+}}f_k(u)$, the hypergeometric equation reads
as
    \bea \label {f49}
         [z(1-z)\frac{\partial^2}{\partial
         z^2}+(2-(\frac{5N-8}{2(N-1)})z)\frac{\partial}{\partial
         z}-\frac{(7-4N)(5-2N)}{16(N-1)^2}]f_k=0,
    \eea
and the solution becomes
   \bea \label {f50}
         S_k(z)=C_{3,k}u^{\alpha_{+}}F(a',b';c';z),
    \eea
where
    \bea \label {f51}
         a'=a=\frac{2N-5}{4(N-1)},\hspace{.5cm}b'=\frac{4N-7}{4(N-1)},\hspace{.5cm}c'=2.
    \eea
The value of $C_{3,k}$ can be fixed by expanding the $S_k$ in the
semiclassical limit in which
    \bea \label {f52}
         S_k=\frac{1}{2\pi i}\oint_{A_k}d
         x{\sqrt{(P_N(x))^2-\Lambda^{4N-4}x^4}}=-\frac{\Lambda^{4N-4}}{2\pi
         i}\oint_{\gamma_k}d x{\frac{x^4}{2P(x)}}+...,
    \eea
where $\gamma_k$ is a counterclockwise around the $k-$th root of
$P_{2N}(x)=x^{2N}-ux^2$
    \bea \label {f53}
         x_k=e^{\frac{\pi i
         k}{N-1}}u^{\frac{1}{2(N-1)}},\hspace{.5cm}k=1,...,2(N-1),
    \eea
so
    \bea \label {f54}
         -\frac{\Lambda^{4N-4}}{2\pi i}\oint_{\gamma_k}d
         x{\frac{x^4}{2P(x)}}=
         -\frac{\Lambda^{4N-4}}{2}\frac{x_k^4}{P'(x_k)}=
         -\frac{\Lambda^{4N-4}}{4(N-1)}\textit{e}^{(\frac{5-2N}{(N-1)})\pi
         i k}u^{\alpha_{+}}.
    \eea
Comparing (\ref {f50}) and (\ref {f54}), one might find
    \be
         C_{3,k}=-\frac{\Lambda^{4N-4}}{4(N-1)}\textit{e}^{(\frac{5-2N}{(N-1)})\pi
         i k},
    \ee
and finally
    \bea \label {f55}
         S_{k}=-\frac{\Lambda^{4N-4}}{4(N-1)}e^{(\frac{5-2N}{(N-1)})\pi
         i k }u^{\alpha_{+}}F(a',b';c';\frac{\Lambda^{2\widehat{h}}}{u^2}).
    \eea
Again, one can obtain a closed relation between $C_{1,k},C_{2,k}$
and $C_{3,k}$ and evaluates the solutions in the regions where
$\mid \frac{u^2}{\Lambda^{4N-4}} \mid<1$.

It is easy to check that the above expression is consistent with
the fact that in the semiclassical limit
    \bea \label {f56}
         \sum_{k=1}^{2N-2}S_k=0.
    \eea
In fact, this relation is a reflection of $\large\textbf{$Z_2$}$
symmetry of SW curve for gauge group $SO(2N)$. We'll get back to
this later.

\subsubsection*{The case $SO(2N+1)$}
In this case
    \bea \label {f57}
         n=2N,\hspace{.5cm}\widehat{h}=2N-1,\hspace{.5cm}l=1,
    \eea
and we again consider that
    \bea \label {f58}
         u_{2N-2}=u,\;\;\;\;\;{\rm other=0}.
    \eea
Now, as we mentioned before since $u_{2N}$ is zero, we can not
derive a generic differential equation for glueball superfield.
So, we focus on the semi classical region where
$\Upsilon\rightarrow 0$ and by changing of variable as
$z=\frac{1}{u}$, the exceptional equation will be changed to
    \bea \label {f59}
         [u^2\partial_u^2+\frac{(N+1)}{(N-1)}u\partial_u+\frac{(2N+1)(3-2N)}{4(N-1)^2}]S_k=0,
    \eea
upon using the series method or changing of variable as $t=lnz$,
one can show that the solutions are
    \bea \label {f61}
         S_k=A_k(\Lambda)u^{\frac{3-2N}{2(N-1)}}+B_k(\Lambda)u^{\frac{2N+1}{2(N-1)}}.
    \eea
In the region where $\Lambda\rightarrow 0$, then $S\rightarrow 0$
and it implies that the solution of the above equation is
$A(\Lambda)u^{\frac{3-2N}{2(N-1)}}$ and in this limit
    \bea \label {f62}
         -\frac{\Lambda^{4N-2}}{2\pi
         i}\oint_{\gamma_k}dx{\frac{x^2}{2P(x)}}=
         -\frac{\Lambda^{4N-2}}{2}\frac{x_k^2}{P'(x_k)},
    \eea
where $x_k=e^{\frac{\pi i k}{N-1}}u^{\frac{1}{2(N-2)}}$ are the
poles of $P_{2N}(x)$. Then it is easy to see that
    \bea \label {f63}
         S_k=-\frac{\Lambda^{4N-2}}{4(N-1)}e^{\frac{3-2N}{(N-1)}\pi i k
         }u^{\frac{3-2N}{2(N-1)}}.
    \eea
Moreover, One can show that in this case too
    \bea \label {f64}
         \sum_{k=1}^{2N-2}S_k=0.
    \eea

\subsection*{The period $S_0$:}
In order to calculate the period $S_0$, we use the Picard-fuchs
equations and obtain the solutions that we described in the
previous sections. But, by a direct calculation, we show that in
the semi classical regions the period $S_0=0$. For this,
considering the case $SO(2N)$ where
    \bea \label {g64}
         S_0=\frac{1}{2\pi i}\oint_{A_{0}}y d x=
         \frac{1}{2\pi i}\oint_{A_0}\sqrt{(x^{2N}-ux^2)^2-\Lambda^{4N-4}x^4} d x,
    \eea
and expanding the above in the region $|\frac{u^2}{
\Lambda^{4N-4}}|>1$ (where the solution $(\ref {f55}$) is valid.),
one can easily show that
    \bea
         S_0=0.
    \eea
Similar arguments can be given for the gauge group $SO(2N+1)$ and
one obtains $S_0=0$ in the semi classical limit. In fact, these
results together with $(\ref {f56})$ are in agreement with the
fact that in the semi classical limit glueball superfield
vanishes.

$\;\;\;\;\;\;$\section{The effective superpotential:}
 Now, we are going to evaluate the expression for effective superpotential.
After the condensation of monopoles, the group structure breaks as
\cite {Elitzur:1996gk,Terashima:1996pn,Fuji:2002vv,Feng:2002gb}
    \bea \label {f65}
         SO(N)\rightarrow SO(\widehat{N}_{0})\times \prod_{k=1}^{n}U(N_k).
    \eea
as before, we consider the case that non of the monopoles become
massless and so
    \bea \label {f66}
         SO(2N)/SO(2N+1)\rightarrow SO(2N_{0})/SO(2N_{0}+1)\times
         \prod_{k=1}^{2N-2}U(1).
    \eea
The general matrix model formula for the effective superpotential
of a $SO(N)$ theory breaking to
$SO(\widehat{N_0})\prod_{k=1}^{n}SO(N_k)$ is \cite
{Fuji:2002vv,Feng:2002gb}
    \bea \label {f67}
         -\frac{1}{2\pi i}{\cal W}_{eff}(S)=
         \frac{1}{2\pi i}2\widehat{N_0}\frac{\partial{{\cal F}}}{\partial{S_0}}+
         \frac{1}{2\pi i}\sum_{k=1}^{n}N_k\frac{\partial{{\cal F}}}{\partial{S_k}}+
         \tau_0\sum_{k=0}^{n}S_k+\sum_{k=0}^{n}b_k S_k,
    \eea
where the constants $b_k$ are integers and are related to periods
as in (\ref {h8}), though in the $SO(N)$ case the definition of
memomorphic one form $T d x$ is different from the $U(N)$ case
(other variables in (\ref {f67}) are the same as in the section
$2$). In this case, if we define the new curve \cite {Feng:2002gb}
    \bea \label {h67}
         z-\frac{2P_{2N}(x,u)}{x^{k}}+\frac{\Lambda^{2\widehat{h}}}{z}=0,
    \eea
then
    \bea \label {h68}
         T d x=\frac{-d z}{z}.
    \eea
Especially, one can show
    \bea \label {h69}
         \frac{1}{2\pi i}\oint_{A_0}T d x=2\widehat{N_0}=2N_0-l.
    \eea
Now, for the curve
    \be
         y^2=(x^{2N}-ux^2)^2-\Lambda^{2\widehat{h}}x^{2k},
    \ee
we calculate these constants.

The constant $N_0$ can be determined by noticing the fact that we
suppose there are't any massless monopoles and so the gauge group
breaks to $SO(2)/SO(3)\times U(1)^{2N-2}$. Moreover, we can
rewrite the
 $P_{2N}(x)=x^{2N_0}\times\prod_{i=1}^{2N-2}(x^2+x_k^2)$ \cite {Fuji:2002vv}
in this case. So, for the above curve $N_0=1$

Furthermore, one may determine the $N_0$ by direct calculation
using the (\ref {h69}) and obtains $N_0=1$. We also obtained that
because of $Z_2$ symmetry
    \bea \label {f71}
         \sum_{k=0}^{2N-2}S_k=0.
    \eea
Again, using the above symmetry, we can obtain a simple expression
for second term as following
    \bea \label {f72}
         \frac{\partial{{\cal F}}}{\partial{S_k}}=\int_{B_k}ydx=
         \lim_{\Lambda_0\rightarrow\infty}[2\int_{x_{k,+}}^{\Lambda_0}
         y d x-2\int_{x_{k,+}}^{\Lambda_0}
         {\cal W} d x-2\Lambda_0{\cal W}(\Lambda_0)],
    \eea
where
    \bea \label {f73}
         x_{k,\pm}=e^{\frac{\pi i k}{N-1}}(u\pm
         \Lambda^{\widehat{h}})^{\frac{1}{2N-2}}.
    \eea
Here $\Lambda_0$ is the point at infinity on the upper sheet.
Considering $x=e^{i\pi}\widetilde{x}$, then
    \bea \label {f74}
         \int_{x_{k,+}}^{\Lambda_0}y d
         x-\int_{x_{k,+}}^{\Lambda_0}{\cal W} d
         x=\int_{x_{k,+}}^{\Lambda_0}\sqrt{(x^{2N}-ux^2)-\Lambda^{2\widehat{h}}x^{2l}d
         x }-\int_{x_{k,+}}^{\Lambda_0}{\cal W}(\Lambda_0)d x \nonumber\\
         =(e^{i\pi})[\int_{x_{k-(N-1),+}}^{\widetilde{\Lambda_0}}
         \sqrt{(\widetilde{x}^{2N}-u\widetilde{x}^2)-\Lambda^{2\widehat{h}}\widetilde{x}^{2l}}d
         \widetilde{x}-\int_{x_{k-(N-1),+}}^{\widetilde{\Lambda_0}}{\cal W}(\widetilde{
         \Lambda_0})d \widetilde{x}],
    \eea
where $\widetilde{\Lambda_0}=\textit{e}^{-i\pi}\Lambda_0$ and we
use
 \bea \label {f76}
         {\cal W}(x)=\frac{1}{2N}x^{2N}-\frac{u}{2}x^2.
    \eea
Then
    \bea \label {f75}
         \frac{\partial{{\cal
         F}}}{\partial{S_k}}=
         e^{i\pi}\frac{\partial{{\cal F}}}{\partial
         S_{k-(N-1)}},
    \eea
Because the gauge symmetry breaks into $U(1)^{2N-2}$ and by a
direct calculation, we have $N_k=1,k=1,...,2N-2$. Hence
    \bea \label {f77}
         \sum_{k=1}^{2N-2}N_k\frac{\partial{{\cal F}}}{\partial{S_k}}
         =0.
    \eea
Now, it remains to determine the $b_k$s which
    \bea \label {f78}
         b_k=-\sum_{j=1}^{k-1}c_j,(b_1=0),\;\;\;\;\;c_j=\frac{1}{2\pi i}\oint_{C_j}T,
    \eea
Note that, since $T_N$ is a logarithmic derivative, its period
integrals are integers. For calculating $c_k$'s, as it has been
showed in \cite {Bertoldi:2003ab}, when $u$ goes to AD points in
$SO(N)$ gauge theory we have
    \bea \label {f80}
         c_k=1,\hspace{.5cm}k=1,...,2N-1.
    \eea
In fact in this limit it can be shown that
    \bea \label {f81}
         c_k=\frac{1}{2\pi i}\oint_{C_k}T_N=\frac{1}{2\pi
         i}\oint_{A_k}T_N=1.
    \eea
In order to calculate the variables we must determine the first
term of ${\cal W}_{eff}$. It is obvious that for $SO(2N)$ gauge
group $2\widehat{N_0}=0$ but for $SO(2N+1)$ case
$2\widehat{N_0}=1$ . So, we should calculate $\frac{\partial {\cal
F}}{\partial S_0}$ for $SO(2N+1)$ gauge group. For this, one
should compute this integral
    \bea \label {h81}
         \frac{\partial {\cal F}}{\partial S_0}=\oint_{B_0}y d
         x=2\int_{0^+}^{\Lambda_0}\sqrt{P_{2N}^2(x)-\Lambda^{2\widehat{h}}x^{2}}.
    \eea
Again, we go to regions where $\Lambda\rightarrow 0$ (as we solved
the picard-fuchs equation in this regions) and rewrite the above
integral around the $\Lambda\rightarrow 0$. Setting $0^+=\Delta$
then
    \bea \label {i82}
         \frac{\partial {\cal F}}{\partial
         S_0}=2\int_{\Delta}^{\Lambda_0}[P_{2N}-\sum_{m=1}^{\infty} \frac{(2m-3)!!}{2m!!}
         \Lambda^{2\widehat{h}m}x^{2-2m}(x^{2N-2}-u)^{1-2m}]d x,
    \eea
and by integrating by part one can show
    \bea \label {j82}
         G(x)&=&\int^x {x'^{2-2m}(x'^{2N-2}-u)^{1-2m}}d x'
         =\hspace{5.3cm}\\
         \nonumber
         &-&\frac{x^{5-2N-2m}(x^{2N-2}-u)^{2-2m}}{(2N-2)(2m-2)}\\
         \nonumber
         &-&\frac{A(2m-1)}{(2N-2)^{2m-1}(2m-2)!}x^{2m(1-2N)+2N+1}
         \ln{(x^{2N-2}-u)}\hspace{3.4cm}\cr
         &-&\sum_{h=1}^{2m-2}\frac{A(h)}{(2N-2)^h}
         \frac{(2m-2-h)!}{(2m-2)!}
         (x)^{2h(1-N)-2m+3}(x^{2N-2}-u)^{h-2m+1}\hspace{1.2cm}\cr
         &+&\frac{A(2m-1)}{(2N-2)^{2m-1}(2m-2)!}
         \int^x x'^{2m-2N(2m-1)}\ln{(x'^{2N-2}-u)}d x',
    \eea
where we suppose that $(x^{2N-2}>u)$ and
$A(k)=\prod_{i=2}^{h}{[2i(1-N)-2m+3]}$. The last integral can be
performed by using the change of variables as $t=\ln(x^{2N-2}-u)$
and defining a function H(x) as
    \bea
         H(x)=(x^{2N-2}-u)(x)^{2m(1-2N)+3},
    \eea
then
    \bea  \label {k82}
         \int^x x'^{2m-2N(2m-1)}\ln{(x'^{2N-2}-u)}d x'=
         \frac{1}{2N-2}\sum_{j=0}^{\infty}\frac{c_j}{(j+1)(j)!}(\ln(x^{2N-2}-u))^{j+1}
    \eea
where $c_j$s are the coefficients of expansion of $H(x)$ around
the $(u+1)^{\frac{1}{2N-2}}$. So
    \bea \label {l82}
         G(x)&=&-\frac{x^{5-2N-2m}(x^{2N-2}-u)^{2-2m}}{(2N-2)(2m-2)}\\
         \nonumber
         &-& \frac{A(2m-1)}{(2N-2)^{2m-1}(2m-2)!}x^{2m(1-2N)+2N+1}
         \ln{(x^{2N-2}-u)}\hspace{3.4cm}\cr
         &+& \frac{A(2m-1)}{(2N-2)^{2m}(2m-2)!}
         \sum_{j=0}^{\infty}\frac{c_j}{(j+1)(j)!}(\ln(x^{2N-2}-u))^{j+1}\cr
         &-& \sum_{h=1}^{2m-2}\frac{A(h)}{(2N-2)^h}
         \frac{(2m-2-h)!}{(2m-2)!}(x)^{2h(1-N)-2m+3}(x^{2N-2}-u)^{h-2m+1}\hspace{1.2cm},
    \eea
and
     \bea \label {m82}
         \frac{\partial {\cal F}}{\partial S_0}=
         &-& \sum_{m=1}^{\infty} \frac{2(2m-3)!!}{2m!!}
         \Lambda^{(4N-2)m}[G(\Lambda_0)-G(\Delta)]\\ \nonumber
         &+& \frac{2}{2N+1}\Lambda_0^{2N+1}-\frac{2}{2N+1}\Delta^{2N+1}-
         \frac{2u}{3}\Lambda_0^3+\frac{2u}{3}\Delta^3.\\ \nonumber
    \eea

After all of these, the effective superpotential is
    \bea \label {n82}
         {\cal W}_{eff}(S)&=&
         -2\widehat{N_0}\frac{\partial{{\cal F}}}{\partial{S_0}}
         -2\pi i\sum_{k=1}^{2N-2}b_k S_k\cr
         &=&-(2N_0-2)\frac{\partial{{\cal F}}}{\partial{S_0}}
         -2\pi i(N-1)\sum_{k=1}^{N-1}S_k,
    \eea
and so, the effective superpotential becomes
    \bea \label {f83}
         {\cal W}_{eff}(S)=\pi i\frac{\Lambda^{4N-4}}{(1-e^{\frac{5-2N}{N-1}\pi
         i})}u^{\frac{5-2N}{2(N-1)}}
         F(\frac{2N-5}{4(N-1)},\frac{4N-7}{4(N-1)},2,
         \frac{\Lambda^{4N-4}}{u^2})\hspace{2cm}
    \eea
for $SO(2N)$, and for $SO(2N+1)$ in the semi classical limit
    \bea \label {f84}
         {\cal
         W}_{eff}(S)&=&
         \pi i\frac{\Lambda^{4N-2}}{(1-e ^{\frac{3-2N}{N-1}\pi
         i})}u^{\frac{3-2N}{2(N-1)}}\\ \nonumber
         &-&\frac{2}{2N+1}(\Lambda_0^{2N+1}-\Delta^{2N+1})
         +\frac{2u}{3}(\Lambda_0^3-\Delta^3)\\ \nonumber
         &+&\sum_{m=1}^{\infty} \frac{2(2m-3)!!}{2m!!}
         \Lambda^{4N-2}[G(\Lambda_0)-G(\Delta)]\hspace{1.3cm}.
    \eea

$\;\;\;\;\;\;$\subsection{Scaling behavior around the AD points}
 In this section, we consider the IR limit $\Lambda\rightarrow\infty$, where the
cycles $A_i$ become large and move out to $\infty$, and do not
effect the IR physics. So, it is enough to study the scaling
behavior of chiral operators around the origin with cycle $A_0$
\cite {Eguchi:2003wv}. It is easy to see that at AD points
    \bea \label {f85}
         U_r=0,\hspace{.5cm}S_r=0,\;\;\; {\rm for all\;\; r}.
    \eea
Perturbing the effective superpotential as
   \bea \label {f86}
         {\cal W}'(x)=x^{2N-1}-u x+
         \sum_{m=0}^{2N-2}g_{2m}x^{2m-1}\Lambda^{2(N-m)},
    \eea
and by computing the expectation value of chiral fields we obtain
the scaling behavior of them. Now, we apply a scale
transformation as
    \bea \label {f87}
         x\rightarrow\rho^\gamma x,\hspace{.5cm}g_{2m}\rightarrow\rho^{\gamma{2(N-m)}}g_{2m}
         ,\hspace{.5cm}
         (m=0,...,2N-2),
    \eea
then
    \bea \label {f88}
         U_r\cong\oint_{A_0}\rho^{\gamma r}x^r\frac{\rho^{2\gamma
         N}(2Nx^{2N-1}-\sum_{2m}x^{2m-1}\Lambda^{2N-2m})} {\rho^{\gamma
         N}(2Nx^{2N-1}-\sum_{2m}x^{2m-1}\Lambda^{2N-2m})}d x,
    \eea
and
    \bea \label {f89}
         U_r\rightarrow\rho^{\gamma(r+N)}U_r.
    \eea
So, the chiral field $U(r)$ has the scaling dimension
$\triangle=\gamma(r+N)$. If we assume that the field Tr$\Phi$ has
no anomalous dimension \cite {Eguchi:2003wv}, then
$\gamma=\frac{1}{N+1}$
 and
    \bea \label {f90}
         \Delta(g_{2m})=2\frac{N-m}{N+1},\hspace{.5cm}\Delta(U_r)=\frac{N+r}{N+1},
         \hspace{.5cm}\Delta(S_r)=\frac{N+r+1}{N+1}.
    \eea
Dimensions of the coupling constants are in agreement with the
known results  \cite {Eguchi:1996vu,Eguchi:1996ds}. Also
    \bea \label {f91}
         \Delta(g_{2m})&\leq &1\;\;\;\Leftrightarrow \;\;\;
         2m{\geq{N-1}}.
    \eea
The case $\Delta(g_{2m})\leq1$ is well-known and should
correspond to a coupling constant in the ${\cal N}=2$
superconformal field theory in $4-$ dimensions \cite
{Argyres:1995xn}.

General chiral perturbation of an ${\cal N}=2$ SCFT$_4$ action
should have the form
    \bea \label {f92}
         \Delta S=\int d^4xd^2\theta^{+}d^2\theta^{-}
         \sum_{m:\Delta (g_{2m})\leq 1}g_{2m}{\cal {O}}_{2m}\\
         \Delta (g_{2m})+\Delta ( {\cal {O}}_{2m})=2,\hspace{1.5cm}
    \eea
so, the chiral operator ${\cal {O}}_{2m}$ is
    \bea \label {f93}
         {\cal{O}}_{2m}=Tr\Phi^{m+2-N},
    \eea
thus, the coupling $g_{2m}$ is paired with $Tr\Phi^{m+2-N}$.

\section {conclusion}

Using the factorization equation of gauge group $SO(N)$ we find
the spectral curve of ${\cal N}=1$ supersymmetric gauge theory and
we obtain a relation between this curve with the vev of monopoles.
Then, using the fact that all monopoles are massive in the system
we obtain general Picard-fuchs equations for classical $U(N)$ and
$SO(N)$ gauge groups respect to parameters of curves that are
casimirs of gauge groups. These differential equations are
hypergeometric equations and have regular singular points which
are dual to Argyres-Douglas points.

Then, we focus on AD points and give the solutions of these
equations and calculate the effective superpotential by using the
geometric picture of $SO(N)$ gauge theory. Scaling behavior in
the IR limits, using the CDSW's machinery, gives us the dimensions
of coupling constants and chiral operators that the results are
consistent with \cite {Eguchi:1996vu}.

\section*{Acknowledgment}

I would like to thank M.Alishahiha for suggesting this problem and
giving me useful comments. I would like to thank A.mosaffa and
S.Parvizi for useful comments. I am obliged to D.Kabirinasab for
helping in editing this paper.


\begin{thebibliography}{99}

\bibitem{Dijkgraaf:2002dh}
R.~Dijkgraaf and C.~Vafa, ``A perturbative window into
non-perturbative physics,'' arXiv:hep-th/0208048.

\bibitem{Dijkgraaf:2002vw}
R.~Dijkgraaf and C.~Vafa, ``On geometry and matrix models,''
Nucl.\ Phys.\ B {\bf 644}, 21 (2002) [arXiv:hep-th/0207106].

\bibitem{Dijkgraaf:2002fc}
R.~Dijkgraaf and C.~Vafa, ``Matrix models, topological strings,
and supersymmetric gauge theories,'' Nucl.\ Phys.\ B {\bf 644}, 3
(2002) [arXiv:hep-th/0206255].

\bibitem{Vafa:2000wi}
C.~Vafa, ``Superstrings and topological strings at large N,'' J.\
Math.\ Phys.\  {\bf 42}, 2798 (2001) [arXiv:hep-th/0008142].

\bibitem{Cachazo:2001jy}
F.~Cachazo, K.~A.~Intriligator and C.~Vafa, ``A large N duality
via a geometric transition,'' Nucl.\ Phys.\ B {\bf 603}, 3 (2001)
[arXiv:hep-th/0103067].

\bibitem{Cachazo:2002pr}
F.~Cachazo and C.~Vafa, ``N = 1 and N = 2 geometry from fluxes,''
arXiv:hep-th/0206017.

\bibitem{Cachazo:2002ry}
F.~Cachazo, M.~R.~Douglas, N.~Seiberg and E.~Witten, ``Chiral
rings and anomalies in supersymmetric gauge theory,'' JHEP {\bf
0212}, 071 (2002) [arXiv:hep-th/0211170].

\bibitem{Seiberg:2002jq}
N.~Seiberg, ``Adding fundamental matter to 'Chiral rings and
anomalies in  supersymmetric gauge theory','' JHEP {\bf 0301},
061 (2003) [arXiv:hep-th/0212225].

\bibitem{Cachazo:2002zk}
F.~Cachazo, N.~Seiberg and E.~Witten, ``Phases of N = 1
supersymmetric gauge theories and matrices,'' JHEP {\bf 0302},
042 (2003) [arXiv:hep-th/0301006].

\bibitem{Cachazo:2003yc}
F.~Cachazo, N.~Seiberg and E.~Witten, ``Chiral Rings and Phases
of Supersymmetric Gauge Theories,'' JHEP {\bf 0304}, 018 (2003)
[arXiv:hep-th/0303207].

\bibitem{Konishi:1983hf}
K.~Konishi, ``Anomalous Supersymmetry Transformation Of Some
Composite Operators In Sqcd,'' Phys.\ Lett.\ B {\bf 135}, 439
(1984).

\bibitem{Konishi:1985tu}
K.~i.~Konishi and K.~i.~Shizuya, ``Functional Integral Approach
To Chiral Anomalies In Supersymmetric Gauge Theories,'' Nuovo
Cim.\ A {\bf 90}, 111 (1985).


\bibitem{Seiberg:1994rs}
N.~Seiberg and E.~Witten, ``Electric - magnetic duality, monopole
condensation, and confinement in N=2 supersymmetric Yang-Mills
theory,'' Nucl.\ Phys.\ B {\bf 426}, 19 (1994) [Erratum-ibid.\ B
{\bf 430}, 485 (1994)] [arXiv:hep-th/9407087].

\bibitem{Seiberg:1994aj}
N.~Seiberg and E.~Witten, ``Monopoles, duality and chiral
symmetry breaking in N=2 supersymmetric QCD,'' Nucl.\ Phys.\ B
{\bf 431}, 484 (1994) [arXiv:hep-th/9408099].

\bibitem{Ferrari:2002jp}
F.~Ferrari, ``On exact superpotentials in confining vacua,''
Nucl.\ Phys.\ B {\bf 648}, 161 (2003) [arXiv:hep-th/0210135].

\bibitem{Ferrari:2002kq}
F.~Ferrari, ``Quantum parameter space and double scaling limits
in N = 1 super  Yang-Mills theory,'' Phys.\ Rev.\ D {\bf 67},
085013 (2003) [arXiv:hep-th/0211069].

\bibitem{Ferrari:2002ad}
F.~Ferrari, ``Non-perturbative double scaling limits,'' Int.\ J.\
Mod.\ Phys.\ A {\bf 18}, 577 (2003) [arXiv:hep-th/0202205].

\bibitem{Ferrari:2002gy}
F.~Ferrari, ``Large N and double scaling limits in two
dimensions,'' JHEP {\bf 0205}, 044 (2002) [arXiv:hep-th/0202002].

\bibitem{Argyres:1995jj}
P.~C.~Argyres and M.~R.~Douglas, ``New phenomena in SU(3)
supersymmetric gauge theory,'' Nucl.\ Phys.\ B {\bf 448}, 93
(1995) [arXiv:hep-th/9505062].

\bibitem{Klemm:1994qs}
A.~Klemm, W.~Lerche, S.~Yankielowicz and S.~Theisen, ``Simple
singularities and N=2 supersymmetric Yang-Mills theory,'' Phys.\
Lett.\ B {\bf 344}, 169 (1995) [arXiv:hep-th/9411048].

\bibitem{Argyres:1994xh}
P.~C.~Argyres and A.~E.~Faraggi, ``The vacuum structure and
spectrum of N=2 supersymmetric SU(n) gauge theory,'' Phys.\ Rev.\
Lett.\  {\bf 74}, 3931 (1995) [arXiv:hep-th/9411057].

\bibitem{Argyres:1995xn}
P.~C.~Argyres, M.~Ronen Plesser, N.~Seiberg and E.~Witten, ``New
N=2 Superconformal Field Theories in Four Dimensions,'' Nucl.\
Phys.\ B {\bf 461}, 71 (1996) [arXiv:hep-th/9511154].

\bibitem{Eguchi:1996vu}
T.~Eguchi, K.~Hori, K.~Ito and S.~K.~Yang, ``Study of $N=2$
Superconformal Field Theories in $4$ Dimensions,'' Nucl.\ Phys.\
B {\bf 471}, 430 (1996) [arXiv:hep-th/9603002].

\bibitem{Eguchi:1996ds}
T.~Eguchi and K.~Hori, ``N = 2 superconformal field theories in 4
dimensions and A-D-E  classification,'' arXiv:hep-th/9607125.

\bibitem{Eguchi:2003wv}
T.~Eguchi and Y.~Sugawara, ``Branches of N = 1 vacua and
Argyres-Douglas points,'' JHEP {\bf 0305}, 063 (2003)
[arXiv:hep-th/0305050].

\bibitem{Bertoldi:2003ab}
G.~Bertoldi, ``Matrix models, Argyres-Douglas singularities and
double scaling limits,'' JHEP {\bf 0306}, 027 (2003)
[arXiv:hep-th/0305058].

\bibitem{Ahn:1997wh}
C.~h.~Ahn, K.~Oh and R.~Tatar, ``M theory fivebrane and confining
phase of N = 1 SO(N(c)) gauge  theories,'' J.\ Geom.\ Phys.\  {\bf
28}, 163 (1998) [arXiv:hep-th/9712005].

\bibitem{Edelstein:2001mw}
J.~D.~Edelstein, K.~Oh and R.~Tatar, ``Orientifold, geometric
transition and large N duality for SO/Sp gauge  theories,'' JHEP
{\bf 0105}, 009 (2001) [arXiv:hep-th/0104037].

\bibitem{Fuji:2002vv}
H.~Fuji and Y.~Ookouchi, ``Confining phase superpotentials for
SO/Sp gauge theories via  geometric transition,'' JHEP {\bf 0302},
028 (2003) [arXiv:hep-th/0205301].

\bibitem{deBoer:1997ap}
J.~de Boer and Y.~Oz, ``Monopole condensation and confining phase
of N = 1 gauge theories via  M-theory fivebrane,'' Nucl.\ Phys.\
B {\bf 511}, 155 (1998) [arXiv:hep-th/9708044].

\bibitem{Ito:1996sq}
K.~Ito and S.~K.~Yang, ``Picard-Fuchs equations and prepotentials
in N = 2 supersymmetric {QCD},'' arXiv:hep-th/9603073.

M.~Alishahiha, ``On the Picard-Fuchs equations of the SW models,''
Phys.\ Lett.\ B {\bf 398}, 100 (1997) [arXiv:hep-th/9609157].

J.~M.~Isidro, A.~Mukherjee, J.~P.~Nunes and H.~J.~Schnitzer, ``A
note on the Picard-Fuchs equations for N = 2 Seiberg-Witten
theories,'' Int.\ J.\ Mod.\ Phys.\ A {\bf 13}, 233 (1998)
[arXiv:hep-th/9703176].

 K.~Ito, ``Picard-Fuchs equations and prepotential in N = 2 supersymmetric G(2)  Yang-Mills theory,''
Phys.\ Lett.\ B {\bf 406}, 54 (1997) [arXiv:hep-th/9703180].

M.~Alishahiha, ``Simple derivation of the Picard-Fuchs equations
for the Seiberg-Witten  models,'' Phys.\ Lett.\ B {\bf 418}, 317
(1998) [arXiv:hep-th/9703186].

Y.~Ohta, ``Picard-Fuchs ordinary differential systems in N = 2
supersymmetric  Yang-Mills theories,'' J.\ Math.\ Phys.\  {\bf
40}, 3211 (1999) [arXiv:hep-th/9812085].

\bibitem {Bateman1955}
Bateman Manuscript Project , "Higher transcendental
functions",Vol.I,New York,McGraw-Hill,1953-55,Arthur Erdeyi
Editor.

\bibitem{Elitzur:1996gk}
S.~Elitzur, A.~Forge, A.~Giveon, K.~A.~Intriligator and
E.~Rabinovici,
Phys.\ Lett.\ B {\bf 379}, 121 (1996) [arXiv:hep-th/9603051].

\bibitem{Feng:2002gb}
B.~Feng, ``Geometric dual and matrix theory for SO/Sp gauge
theories,'' arXiv:hep-th/0212010.

\bibitem{Terashima:1996pn}
S.~Terashima and S.~K.~Yang, ``Confining phase of N = 1
supersymmetric gauge theories and N = 2  massless solitons,''
Phys.\ Lett.\ B {\bf 391}, 107 (1997) [arXiv:hep-th/9607151].

S.~Terashima and S.~K.~Yang,
Nucl.\ Phys.\ B {\bf 519}, 453 (1998) [arXiv:hep-th/9706076].

S.~Terashima,
Nucl.\ Phys.\ B {\bf 526}, 163 (1998) [arXiv:hep-th/9712172].









\end{thebibliography}
\end{document}